# Estimation of direct laser acceleration in laser wakefield accelerators using particle-in-cell simulations

**J L Shaw[1], N Lemos[1], K A Marsh[1], F S Tsung[2], W B Mori[1,2] and C Joshi[1]**

[1]Department of Electrical Engineering & [2]Department of Physics and Astronomy, University of California Los Angeles, 405 Hilgard Ave, Los Angeles, California, United States 90095

Email: jshaw05@ucla.edu

**Abstract.** Many current laser wakefield acceleration (LWFA) experiments are carried out in a regime where the laser pulse length is on the order of or longer than the wake wavelength and where ionization injection is employed to inject electrons into the wake. In these experiments, the trapped electrons will co-propagate with the longitudinal wakefield and the transverse laser field. In this scenario, the electrons can gain a significant amount of energy from both the direct laser acceleration (DLA) mechanism as well as the usual LWFA mechanism. Particle-in-cell (PIC) codes are frequently used to discern the relative contribution of these two mechanisms. However, if the longitudinal resolution used in the PIC simulations is inadequate, it can produce numerical heating that can overestimate the transverse motion, which is important in determining the energy gain due to DLA. We have therefore carried out a systematic study of this LWFA regime by varying the longitudinal resolution of PIC simulations from the standard, best-practice resolution of 30 points per laser wavelength to four times that value and then examining the energy gain characteristics of both the highest-energy electrons and the bulk electrons. We show that for a resolution of between 60 and 120 points per laser wavelength, the final energy spectra of the accelerated electrons converge in spectral shape. By calculating the contribution of DLA to the final energies of the electrons produced from the LWFA, we find that although the transverse momentum and oscillation radii are over-estimated in the lower-resolution simulations, this over-estimation does not lead to artificial energy gain by DLA. Rather, the DLA contribution to the highest-energy electrons is larger in the higher-resolution cases because the DLA resonance is better maintained. Thus, even at the highest longitudinal resolutions, DLA contributes a significant portion of the energy gained by the highest-energy electrons and also contributes to accelerating the bulk of the charge in the electron beam produced by the LWFA.

## 1. Introduction

Laser wakefield acceleration (LWFA) [1] is being pursued as a concept for building compact accelerators. One of the promising LWFA regimes for producing high-energy and high-quality electron beams is known as the blowout regime [2]. In the ideal blowout regime, an intense ($a_0 \sim 2 - 4$), fs-scale ($c\tau_p < R_b$) laser pulse matched to the plasma density ($w_0 \sim R_b$) is propagated through an underdense plasma. Here, $a_0 = \frac{eA}{mc^2}$ is the normalized vector potential, A is the vector potential, e is the charge of the electron, m is the mass of the electron, c is the speed of light, $\tau_p$ is the pulse length of the drive laser, $R_b = \frac{2\sqrt{a_0}c}{\omega_p}$ is the blowout radius, $\omega_p = \sqrt{\frac{e^2 n_e}{m\epsilon_0}}$ is the plasma frequency, $n_e$ is the plasma density, $\epsilon_0$ is the permittivity of free space, and $w_0$ is the laser spot size. Typically, the front portion of this laser pulse ionizes a gas via tunnel ionization [3], and then the pondermotive force of the laser expels those plasma electrons. On this femtosecond timescale, the more massive ions are relatively immobile, so an ion column forms behind the laser pulse. The expelled plasma electrons are drawn back towards the laser axis by the Coulomb force of the ion column where they overshoot the axis and set up a wake structure. Electrons can become trapped in this wake structure when their velocity reaches the phase velocity of the wake and can then be accelerated to relativistic energies by the longitudinal electric field established by the wake.

While few recent LWFA experiments have used laser pulse lengths and intensities required to operate in the blowout regime, far more experiments are still carried out using drive laser pulse lengths that are on the order of the wake wavelength or longer. Furthermore, ionization injection [4, 5] is often employed to inject charge to be accelerated into the wake using laser pulses that are less intense than required for self-trapping of the plasma electrons. In these experiments, there is the possibility of an overlap of the injected electrons and the laser field. Several studies are beginning to study this hybrid regime where the longitudinal wakefield (LWFA) and the transverse laser field (often referred to as direct laser acceleration (DLA) [6, 7]) can both contribute significantly to the ultimate energy gain, divergence, and energy spectrum of the accelerated electrons. Many of the experimental signatures that might give a clue as to the relative importance of these two mechanism such as the transverse ellipticity and the highest energy of the accelerated bunch can be predicted using particle-in-cell (PIC) code simulations. However, it is important to be certain that the conclusions are physical and not due to numerical artefacts. It is the purpose of this paper to study how the longitudinal resolution used in a PIC code can affect the interpretation of the relative roles of LWFA and DLA in the energy gained by the electrons in many current experiments and to determine the contribution of DLA to the overall accelerated spectrum.

## 2. Background

It was recently observed [8] in PIC code simulations of a LWFA operating in a near-blowout regime that if the pulse length of the drive laser is such that it overlaps the trapped electrons, a second acceleration mechanism known as direct laser acceleration (DLA) can be induced in addition to the wake acceleration. The wake structure in a LWFA is ideal for DLA because as the trapped electrons are accelerated by the wake, they execute betatron oscillations [9-11] in the ion column formed behind the drive laser pulse. Therefore, when the drive pulse overlaps the trapped electrons, then those electrons that undergo betatron oscillation in the plane of the laser can be accelerated by the transverse electric field of the laser. This increased transverse momentum can then be converted to longitudinal momentum via the $\vec{v} \times \vec{B}$ force of the laser. This process, which is essentially the inverse of the ion channel laser process [12] and is closely related to the inverse free electron laser acceleration mechanism [13, 14], is described by the one-dimensional resonance condition $N\omega_\beta = \left(1 - \frac{v_\parallel}{v_\phi}\right)\omega_0$ [6] where N is an integer number, $\omega_\beta = \frac{\omega_p}{\sqrt{2\gamma}}$ is the betatron frequency, $\gamma$ is the Lorentz factor of the electron, $v_\parallel$ is the velocity of the electron in the longitudinal direction, $v_\Phi$ is the phase velocity of the laser, and $\omega_0$ is the laser frequency.

It seems unlikely that this resonance condition could be satisfied in a LWFA since the betatron frequency and longitudinal velocity of the trapped electrons change continuously as they are accelerated in the wake and the laser frequency and phase velocity evolve over the propagation distance. However, electrons in a LWFA can gain sizeable energy from DLA even if they only intermittently satisfy the DLA resonance condition [15]. When DLA contributes in this manner, it can contribute significantly to the final energy of the highest-energy electrons, and its contribution to the final electron energy scales with the degree of overlap between the laser and trapped electrons [8]. Furthermore, for a LWFA operating in either the forced [16] or self-modulated LWFA [17]/ Raman forward scattering [18, 19] regime, the DLA contribution to the final energy of the electrons can exceed the energy contribution due to the wakefield [8]. The idea of using DLA to increase energy gain in LWFAs was further extended in simulation by Zhang et al. [20], who proposed a configuration that uses one laser pulse to drive the wake and a second laser pulse that overlaps with the trapped electrons.

In 2014, Lehe et al. [21] found that in PIC code simulations of a laser field interacting with an electron in vacuum, numerical effects can lead to an overestimation of the transverse momentum and excursion of an electron and can ultimately lead to spurious heating. This overestimation can happen when the magnetic field is not calculated with enough precision in a PIC code that is using the standard Yee algorithm [22]. In such a code, the macroparticles are distributed across the grid and therefore may not fall exactly on the grid itself. Consequently, before calculating the force on each particle, the electric and magnetic fields need to be extrapolated from the grid to the location of the macroparticles. Additionally, since the magnetic field is defined at half time steps in the Yee lattice, it needs to be interpolated in time since the Lorentz force on the macroparticles is calculated at integer time steps. If a simple time average is used to interpolate the magnetic field, the resulting error can lead to artificial growth in the transverse trajectories and momentum of the electrons; this artificial growth disappears when a third-order interpolation method is used [21]. Since the DLA mechanism in References [8] and [20] depends upon the transverse motion of the electrons in an ion column, in this paper we explore the extent to which the DLA of electrons in LWFAs may be affected by these numerical effects in PIC code simulations. We find that contrary to expectation, the DLA effect persists and can significantly contribute to the total energy of the electrons even when the simulation resolution is significantly increased.

## 3. Simulation Setup

To explore if the numerical effects are influencing the estimation of the relative contribution of DLA in LWFA, we ran a series of OSIRIS [23] PIC code simulations with increasing longitudinal and transverse resolutions. In the transverse direction, the laser phase is approximately constant and the laser amplitude envelope is slowly varying on the scale of the grid, so the transverse resolution is not expected to have a large impact on the results. Indeed, there was no appreciable difference when the transverse resolution was increased, and therefore only the effects of the longitudinal resolution are presented in this paper. The longitudinal resolution is expected to have a much larger effect because the laser phase varies rapidly in the longitudinal direction and the speed of light in simulations is dominated by the value of the longitudinal resolution. The longitudinal resolution effects were examined by reducing the grid size in the longitudinal direction. This increase in longitudinal resolution increases the accuracy when extrapolating the fields to the macroparticles, and subsequently reduces the time step, which improves the accuracy of the interpolation of the magnetic field in time. Though the best practices in the choice of resolutions used in PIC code simulations of LWFA vary from research group to research group, our research group at UCLA has settled on a standard resolution of 40 grid points per laser spot size and of 30 grid points per laser wavelength (pts/$\lambda_0$) [24] for LWFA simulations. Testing via convergence studies [24] has demonstrated that when using this longitudinal resolution in OSIRIS, any error in the laser propagation becomes small and the wakefields converge.

The laser and plasma parameters used in this particular series of simulations model the achievable experimental parameters of the ongoing experiments on the 20 TW Ti:Sapphire laser in the XPL lab at UCLA. The simulation box is comprised of a 63.63 μm by 18.86 μm window that moves at the speed of light. The grid in that window is 2342 x 472. The laser pulse is focused to a spot size $w_0$ of 7.2 μm halfway up a 150-μm-long density up-ramp. The laser pulse ionizes an initially-neutral mixture of 0.5% $N_2$ and 99.5% He so that charge is trapped in the wake via ionization injection [4, 5]. The produced plasma density has a maximum value of 6.3 x $10^{18}$ $cm^{-3}$ and a constant-density plateau of 1400 μm with 150-μm-long linear density up- and down-ramps. The laser pulse has a central wavelength of 815 nm and an $a_0$ of 2.8. Its pulse length is 60 fs (FWHM of intensity) so that the laser overlaps the trapped electrons; thus DLA is expected to contribute to the final energy of these electrons. The standard-resolution simulation uses $k_0\Delta z = 0.209$ and $k_p\Delta x = 0.085$ to resolve the longitudinal and transverse dimensions, respectively. The resulting normalized time step is 0.01265.

The next two simulations in the series are identical to the standard-resolution simulation with the exception of the number of grid points in the longitudinal direction and the subsequent change in the time step. For the double-resolution case (where the longitudinal resolution is increased to 60 points per laser wavelength), the grid is 4682 x 472. The resulting $k_0\Delta z$ is reduced to 0.105, and the normalized time step is reduced to 0.00638. For the quadrupole-resolution case (longitudinal resolution of 120 points per laser wavelength), the grid is 9368 x 472. The resulting $k_0\Delta z$ is further reduced to 0.052 and results in a reduced normalized time step of 0.00320. Each simulation was run to completion, and the trapped electrons were filtered to select only those electrons that were accelerated in the first bucket and exited the plasma. From those electrons, the 5 highest-energy electrons and 500 random electrons were tagged. The simulation was then rerun with those tagged particles so that the momentum, position, and fields sampled by those electrons at each time step are known. From this information, it is possible to determine whether the electrons gain energy from DLA, from LWFA, or from both.

## 4. Results

When DLA is present in a LWFA, the electrons' motions become remarkably different than in the case with only wakefield acceleration. This difference is especially pronounced when ionization injection is used because the electrons are ionized further back inside the first period of the wake and they can gain sizeable transverse momentum as they slip backwards in the wake. These two effects can change the position of the electron trapping relative to the case where no laser field is present. Also, the effective distance for dephasing in the longitudinal field is affected by both this change in the trapping location and the length of the laser pulse relative to the wake wavelength. Furthermore, electrons can continue to gain energy due to DLA even after dephasing with respect to the wake. In this case, the final energy of the electrons may increase or decrease depending on the relative contribution of the two mechanisms to the overall energy of the particle. To isolate the LWFA effect, we therefore carried out one additional simulation where the pulse length was reduced to 35 fs so that there is no overlap between the laser and the trapped electrons. This additional simulation is identical to the standard-resolution simulation in all other respects and therefore the results serve as a baseline to examine the transverse momentum and oscillation when DLA is not present.

The oscillation radius and the transverse momentum of 5 of the highest-energy $W_{max}$ electrons for the simulations with a 35 fs pulse (where no DLA is present and $W_{max}$ = 331 MeV) and a 60 fs pulse (where DLA contributes to the energy gained by the electrons and $W_{max}$= 199 MeV) are compared in figure 1. Figure 1a shows that the typical extent of the betatron oscillation in the 60 fs case is ~ 4 μm, which is more than 5 times the typical excursion of ~ 0.7 μm in the 35 fs simulation. Furthermore, figure 1b shows that the normalized transverse momentum $p_\perp/mc$ can reach 35 when DLA is present in the $\tau_p$ = 60 fs case, where the wake-only $\tau_p$ = 35 fs case reaches a much more modest 5. It is clear from this figure that both the transverse oscillation amplitude (figure 1a) and momentum (figure 1b) are significantly increased when the laser pulse overlaps the trapped electrons and DLA is present in addition to LWFA, and it emphasizes why it is critical to determine whether these dramatic increases in the transverse momentum and excursion are physical or merely a numerical artefact of the PIC algorithm as described earlier.

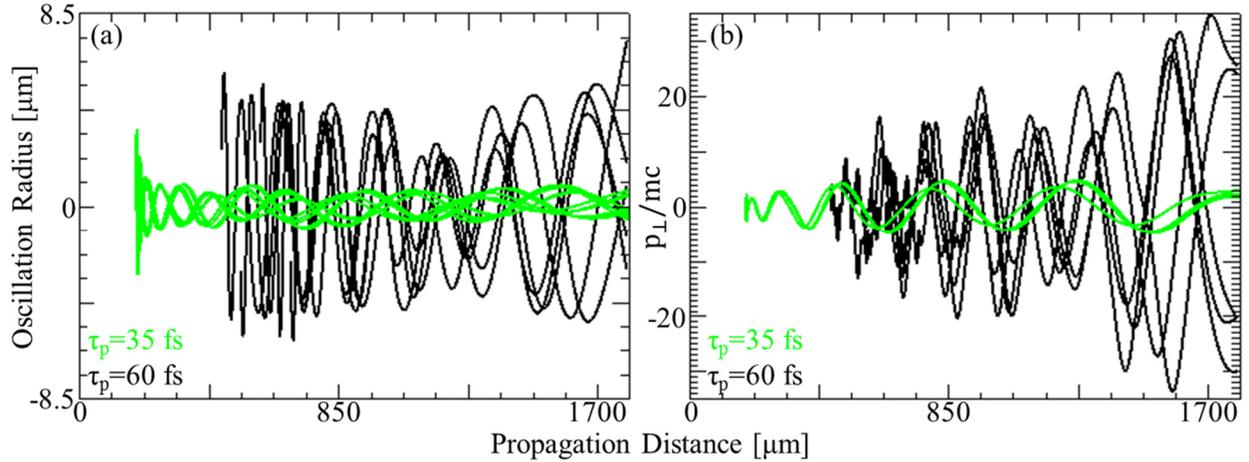

Figure 1: (a) Plot of the electron trajectories for the 5 highest-energy electrons in the $\tau_p = 35$ fs (green curves) and $\tau_p = 60$ fs (black curves) simulations. (b) Normalized transverse momentum as a function of the propagation distance for the $\tau_p = 35$ fs (green curves) and $\tau_p = 60$ fs (black curves) simulations. Note that the highest-energy electrons are trapped much earlier in the $\tau_p = 35$ fs case.

To investigate if a reduced resolution can lead to an artificial growth of the transverse oscillation extent of electrons in a laser field [21], we carried out the exact same 60 fs simulation but with three different longitudinal resolutions (see figure 2). The color bar has been slightly saturated to emphasize the extent of the trapped charge. For the standard-resolution case (figure 2a), the trapped charge has a large transverse extent of ~ ±8.5 µm and is deeply modulated at half of the laser wavelength [25]. Here, the charge density is the largest at the farthest extents of the modulation. When the resolution is doubled (figure 2b), the bunching is reduced. Though there is still charge bunched at half of the laser wavelength, its maximum transverse extent is reduced to ~ ±7 µm and the depth of modulation has decreased. In this case, more of the trapped charge is located on-axis relative to the standard-resolution case. The bunching is further washed out when the resolution is quadrupled (figure 2c). This finding suggests that reduced resolution indeed enhances the transverse motion of the electrons.

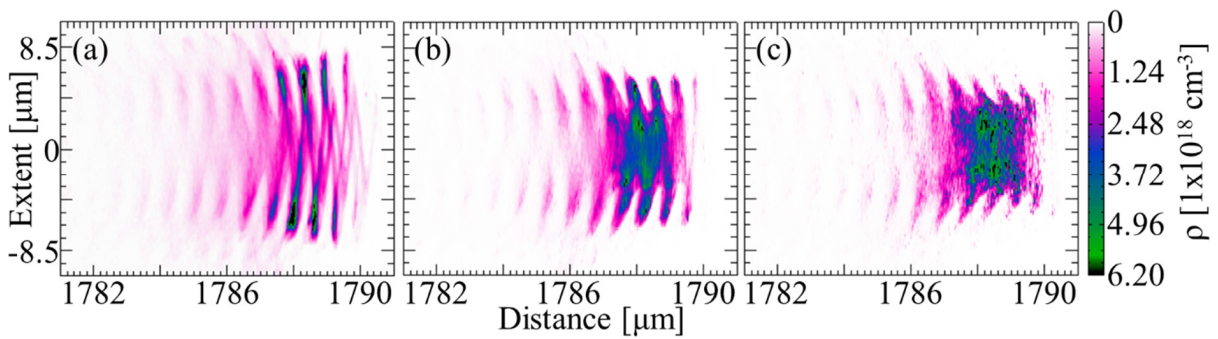

Figure 2: Plots of the trapped charge after it exits the plasma for the (a) standard-, (b) double-, and (c) quadrupole-resolution simulations. $\tau_p = 60$ fs in all cases.

In a LWFA where the laser pulse overlaps the trapped electrons, DLA can be expected to contribute sizably to the final energy of the highest-energy electrons [8, 20]. If this energy gain due to DLA is purely a numerical artefact, then the reduced extent of the transverse oscillations seen in the higher-resolution simulations in figure 2 should correlate with a decreased DLA contribution to the energy gained by the electrons and hence a reduction in the maximum electron energies produced by the LWFA. However, as figure 3 shows, the high-energy tails of the electron spectra in the higher-resolution simulations actually push to larger energies, and the electron spectra converge within 10% for resolutions greater than 60 grid points per laser wavelength due to the improved interpolation of the magnetic field and the improved extrapolation to the macroparticles. To understand why the higher-energy tails of the spectra shown in figure 3 do not decrease with resolution as expected given the results of figure 2, the relative contributions to the final energies of those highest-energy electrons must be explored.

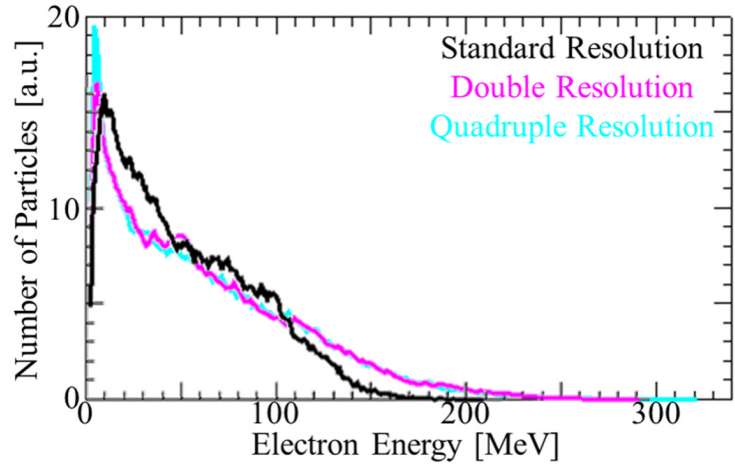

Figure 3: Final energy spectra of the electrons accelerated in the first bucket of the wake for the standard 30 pts/$\lambda_0$ (black), double 60 pts/$\lambda_0$ (magenta), and quadrupole 120 pts/$\lambda_0$ resolution (cyan) simulations. The two higher-resolution spectra have converged. The small random variation between the three spectra arises because only 20% of the macroparticles are extracted and saved.

Using the values from tracking the 5 highest-energy particles, the contribution to the final energy of the electrons due to the longitudinal fields $\vec{E}_\parallel$ can be calculated for each time step of the simulation via

$$W_\parallel = \int_0^t \vec{E}_\parallel \cdot \vec{v}_\parallel dt' \quad (1)$$

where $\vec{v}_\parallel$ is the velocity of the electron in the longitudinal direction. The dominant longitudinal field is the wakefield, and therefore this value will be called the "LWFA Contribution" to the final electron energy. Similarly, the contribution to the final electron energy due to the transverse fields $\vec{E}_\perp$ is given by

$$W_\perp = \int_0^t \vec{E}_\perp \cdot \vec{v}_\perp dt' \quad (2)$$

In this equation, $\vec{v}_\perp$ is the velocity of the tracked electron in the plane of the laser polarization. The dominant transverse field is the transverse laser field, and so this value will be called the "DLA contribution" to the final electron energy.

Figure 4a shows that in the standard-resolution $\tau_p = 60$ fs simulation, the trapped electrons reach a maximum energy of 199 MeV upon exiting the plasma. Comparing the LWFA contribution (red curve) to the total electron energy curve (black curve) shows that the wakefield only accounts for approximately half of the total energy gained by the electrons. The other half of the energy gains comes from DLA as shown by the blue curve in figure 4a. However, these electrons do not continuously gain energy from

DLA. Rather, once they are trapped, they gain energy from DLA for ~ 360 μm before coming out of resonance and losing all the energy that they had gained from DLA. At ~ 1250 μm, they regain resonance and are accelerated by DLA until exiting the plasma.

A similar comparison for the double-resolution case (figure 4b) shows that the DLA contribution to the final energy gain is not merely a numerical artefact. In this case, the maximum energy reaches 285 MeV, and the DLA contribution has actually increased relative to the standard-resolution case because the electrons are able to maintain the DLA resonance condition for much longer distances. In this case, after the electrons are trapped, they gain a small amount of energy from DLA and then quickly lose it. But then they regain the DLA resonance and maintain it for a long acceleration distance, which permits large energy gain from DLA. In fact, for the highest-energy electrons in the double-resolution case, LWFA only contributes up to 31.6 MeV to the final electron energy, and some electrons are actually losing energy to the wake as they exit the plasma. The LWFA contribution is reduced in this case relative to the standard-resolution case because these electrons propagate farther forward in the wake and are therefore in either a weakly-accelerating or a decelerating longitudinal field as they continue to gain energy from DLA.

In the quadruple-resolution case (figure 4c), the maximum electron energy has increased yet again to a final energy of 328 MeV at the exit of the plasma. The DLA contribution to the highest-energy electrons has increased 28% relative to the double-resolution case, and all of the 5 highest-energy electrons are actually losing energy to the wake. In this case, the DLA contribution has increased even relative to the double-resolution case because the electrons come into resonance earlier and therefore gain energy from DLA for an even longer acceleration length. Further, the gradient is 10% larger in this case, which coupled with the larger loss of energy to the wake, means that the electrons have dephased and are propagating farther forward in the wake relative to the double-resolution case in a decelerating longitudinal field while continuing to be accelerated by DLA in a higher-intensity portion of the laser pulse.

The differing ability of each resolution case to maintain DLA resonance can be clearly seen when examining the trajectories of one of the 5 highest-energy electrons as shown in row 2 of figure 4. The oscillation radius is expected to be largest for the standard-resolution case (figure 4d), and indeed we see that when the electron is first trapped and gaining energy from DLA, it has the largest radius. However, as soon as that electron comes out of resonance, its radius is drastically reduced. Yet once it regains resonance, its transverse oscillation amplitude quickly surpasses the oscillation radii of the double- (figure 4e) and quadruple- (figure 4f) resolution cases. In the double-resolution case, when the electron is initially in resonance, its oscillations grow in radii, but not to as large of an extent as in the standard-resolution case. The electron then comes out of resonance and the oscillation radius decreases. It then regains resonance and the oscillation radii again grow, but again these amplitudes are less than in the standard-resolution case. In the quadruple-resolution case (figure 4f), the electron is trapped and then comes in and out of the DLA resonance in a short distance. Then that electron again comes into DLA resonance, and its oscillation radius grows slowly as the electron remains in resonance. However, even with the sustained resonance, the oscillation radii never grow larger than in the standard-resolution case. Therefore, looking at the trajectories supports the idea that lower resolutions can lead to enhanced transverse motion, but it also shows that such an increase does not lead to artificial energy gain by DLA because the electrons cannot maintain the DLA resonance condition in lower-resolution simulations.

It is also predicted that the transverse momentum should be artificially enhanced in the lower-resolution case. Examining the normalized transverse momentum of the electrons (third row of figure 4), shows the same trend as the transverse oscillations— the transverse momentum is reduced when an electron is out of DLA resonance, yet once in DLA resonance, the transverse momentum of the electron in the standard-resolution case quickly grows to larger values than in the double- and quadruple-resolution cases.

Thus, figure 4 confirms that DLA is not purely a numerical artefact but rather that it is a physical effect that continues to play a large role in the energy gained by the highest-energy electrons produced from a LWFA even at increased resolutions. Furthermore, figure 4 shows that the highest electron energies actually increase with increasing resolution because the electrons are better able to maintain the DLA resonance. The resulting increase in the DLA contribution to the final electron energies explains why the maximum energy of the electron spectra (figure 3) increases with resolution despite the reduction in the transverse extent as seen in figure 2.

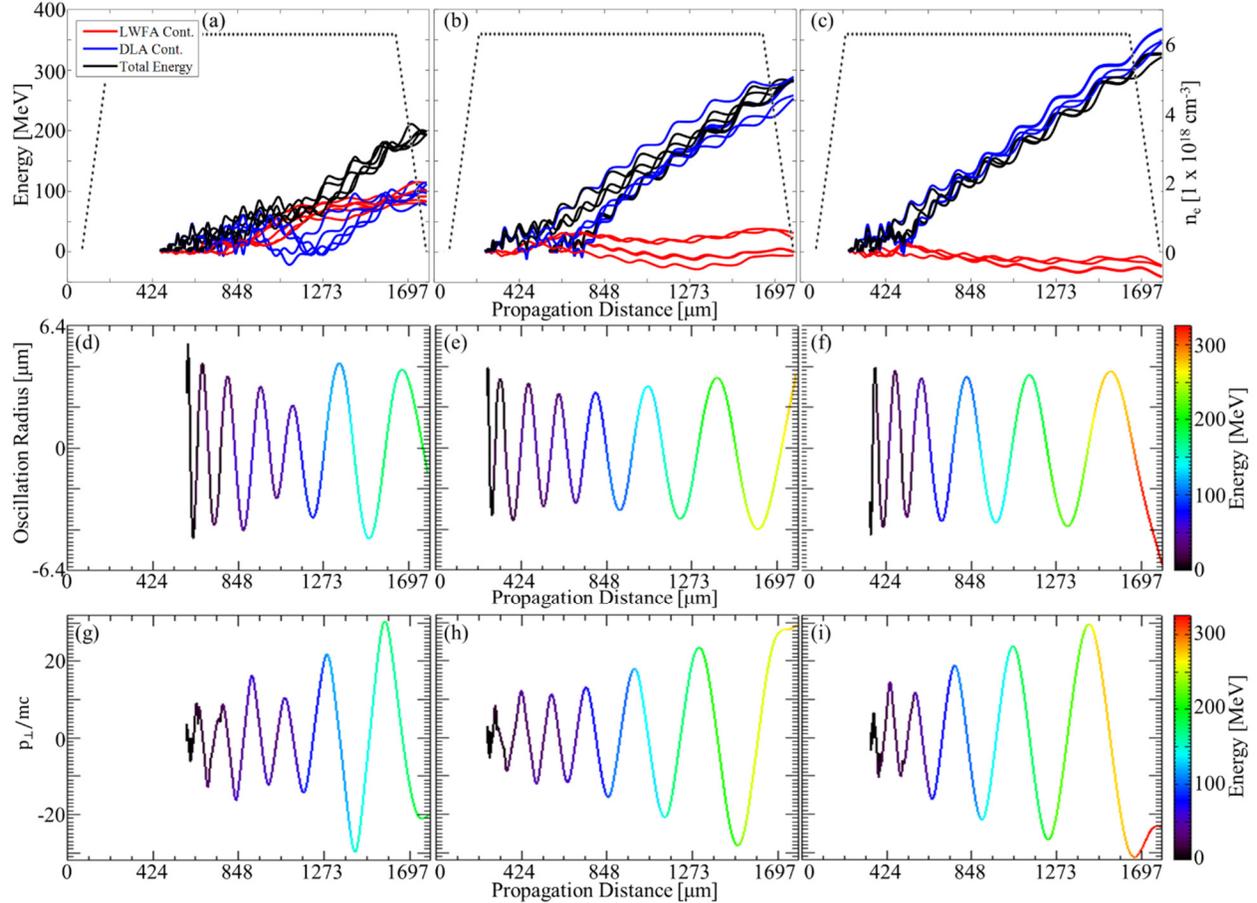

Figure 4: First Row: Plots of total electron energy (black curves), LWFA contribution to the electron energy (red curves), and DLA contribution to the electron energy (blue curves) as a function of the propagation distance for the (a) standard-, (b) double-, and (c) quadruple-resolution cases. Curves are shown for the 5 highest-energy electrons in each simulation. Black dotted line shows the plasma density profile used in the simulation. For all three cases, there was ~ 10% residual electron density on axis inside the first period of the wake. Second Row: Plots of the trajectories of one of the 5 highest-energy electrons for the (d) standard-, (e) double-, and (f) quadruple-resolution cases. Third Row: Plots of the normalized transverse momentum of the same electrons as in row 2 as a function of propagation distance for the (g) standard-, (h) double-, and (i) quadruple-resolution simulations.

The relative contribution of DLA and LWFA to the final energies of the 500 randomly-tagged electrons is next examined to determine if the role that DLA plays in accelerating the bulk of the electrons also increases when the resolution is increased. The first row in figure 5 shows the histograms of the final electron energies for the 500 randomly-tagged electrons for the three different longitudinal resolution simulations. Like the electron spectra for all the electrons in the first bucket of the wake that was shown in figure 3, for the 500 randomly-tagged particles shown here, the energies of the highest-energy electrons actually increase with increasing resolution.

For each resolution, the LWFA contribution to the final electron energy was calculated for the 500 random electrons using (1) and is shown in row 2 of figure 5. For the standard-resolution case (figure 5d), 92% of the electrons gain some energy from LWFA. For the double- (figure 5e) and quadruple- (figure 5f) resolution cases, this percentage is 81% and 87%, respectively. These percentages are similar and shows that LWFA is still important in accelerating the bulk of the charge in a LWFA where the drive laser overlaps the trapped electrons. In the double- and quadruple-resolution cases, the maximum LWFA contributions of 200 MeV and 175 MeV, respectively, cannot account for the maximum energies of 275 MeV for the bulk electrons. Therefore, as already seen in figure 4, DLA is necessary in addition to LWFA to reach those highest electron energies.

The third row of figure 5 shows the histograms of the DLA contribution to the final electron energy of the randomly-tagged particles as calculated using (2). For the standard-resolution case (figure 5g), the maximum DLA contribution is 100 MeV, and 87% of the electrons gain some energy from the DLA mechanism. In the double-resolution case (figure 5h), the maximum DLA contribution is 250 MeV, and 76% of the electrons gain energy from DLA. In the quadruple-resolution case, the maximum DLA contribution is 250 MeV, and 83% of the electrons gain energy from DLA. These percentages are similar in all three cases and show that DLA is also important for accelerating the bulk of the electrons. Furthermore, even in the bulk of the electrons, the maximum DLA contribution increases relative to the standard-resolution simulation when the resolution is increased. Furthermore, the shapes of the DLA histograms show that the distribution of electron energies shifts towards higher energies with increasing resolution, so more electrons have larger DLA contributions to their final energies.

Row 4 of figure 5 shows the histograms of the percent DLA contribution to the final electron energy. The percent DLA contribution is calculated by finding what percentage of an electron's final energy came from DLA. A value of 100% means that all of the electron's final energy can be accounted for by DLA. Values over 100% mean that the electron actually gained more energy from DLA than its final energy. This process can happen when an electron loses energy to the wake while gaining energy from DLA. For the standard-resolution case, 66% of the electrons gain more than half of their final energy from DLA. For the double- and quadruple-resolution cases, 48% and 37%, respectively, of the electrons gain more than half of their energy from DLA. Clearly, there is a trend towards convergence of not only the shape of the overall spectrum (figure 3), but also of the fraction of the electrons that gain half of their energy from DLA. Based on its contribution to the energy gained by the bulk of the electrons in all three resolution cases, it is clear that DLA plays a significant role.

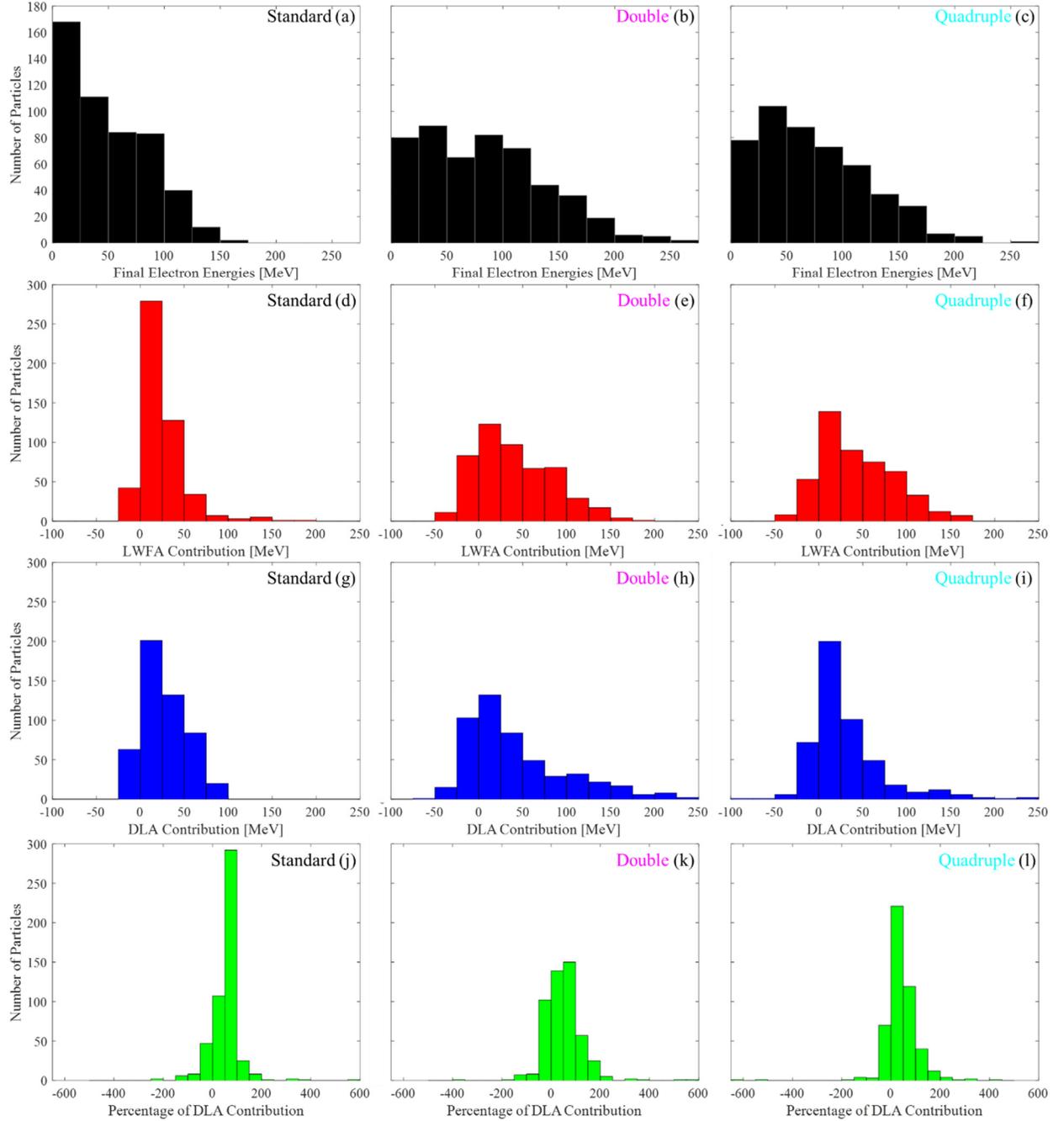

Figure 5: First Row: Histograms of the final electron spectra for the 500 randomly-tagged particles for the (a) standard-, (b) double-, and (c) quadruple-resolution simulations. Second Row: Histograms of the LWFA contribution to the final electron energy for the 500 randomly-tagged particles for the (d) standard-, (e) double-, and (f) quadruple-resolution simulations. Third Row: Histograms of the DLA contribution to the final electron energy for the 500 randomly-tagged particles for the (g) standard-, (h) double-, and (i) quadruple-resolution simulations. Fourth Row: Histograms of the percent of the final electron energy due to DLA for the 500 randomly-tagged particles for the (j) standard-, (k) double-, and (l) quadruple-resolution simulations.

## 5. Conclusion

By investigating a series of OSIRIS PIC code simulations with increasing longitudinal resolutions, this study has demonstrated that the DLA of electrons in a LWFA is not a numerical artefact. Although the transverse momentum and oscillation extent experience some numerical heating in the lower-resolution simulations, that artificial heating does not translate to artificial DLA. In fact, at lower resolutions, the highest-energy electrons are not able to maintain the DLA resonance, which results in lower contributions of DLA to their final energies relative to the higher-resolution simulations. For all resolutions, DLA contributes to the acceleration of the bulk of the electrons, but at the higher resolutions, a larger number of electrons have higher DLA contribution to their final energies.

**Acknowledgements**

Work supported by DOE grant DE-SC0010064 and NSF Graduate Fellowship DGE-0707424. Simulation work done on the Hoffman2 cluster at UCLA and on the Edison cluster at NERSC.